\documentclass[aps,prl,preprint,groupedaddress]{revtex4}
\pdfoutput=1
\usepackage{graphicx}  
\usepackage{dcolumn}   
\usepackage{bm}        
\usepackage{amssymb}   
\usepackage{feynmf}
\usepackage{slashed}
\usepackage{multirow}
\usepackage{mathrsfs}
\def\gsim{\lower0.5ex\hbox{$\:\buildrel >\over\sim\:$}}
\def\lsim{\lower0.5ex\hbox{$\:\buildrel <\over\sim\:$}}

\newcommand{\met}{E_{\rm T}^{\rm miss}}
\newcommand{\be}{\begin{equation}}
\newcommand{\ee}{\end{equation}}
\newcommand{\bea}{\begin{eqnarray}}
\newcommand{\eea}{\end{eqnarray}}

\newcommand{\nbox}{{\,\lower0.9pt\vbox{\hrule \hbox{\vrule height 0.2 cm
\hskip 0.2 cm \vrule height 0.2 cm}\hrule}\,}}

\def\sub#1{_{\lower.25ex\hbox{$\scriptstyle#1$}}}

\newskip\zatskip \zatskip=0pt plus0pt minus0pt
\def\matth{\mathsurround=0pt}
\def\lsim{\mathrel{\mathpalette\atversim<}}
\def\gsim{\mathrel{\mathpalette\atversim>}}
\def\sigv{\ifmmode \langle\sigma v\rangle\else $\langle\sigma v\rangle$\fi}
\newskip\zatskip \zatskip=0pt plus0pt minus0pt
\def\matth{\mathsurround=0pt}
\def\lsim{\mathrel{\mathpalette\atversim<}}
\def\gsim{\mathrel{\mathpalette\atversim>}}
\def\atversim#1#2{\lower0.7ex\vbox{\baselineskip\zatskip\lineskip\zatskip
  \lineskiplimit
  0pt\ialign{$\matth#1\hfil##\hfil$\crcr#2\crcr\sim\crcr}}}

\begin{document}

\vspace{0.5in}

\title{Increasing the Discovery Potential Using Rare SUSY Scenarios Part
I: Gluinos}

\author{Linda M. Carpenter and Khalida Hendricks}
\affiliation{Department of Physics and  \\
   Center for Cosmology and AstroParticle Physics, CCAPP \\
   Ohio State University 191 W Woodruff Ave \\
   Columbus, OH U.S.A. \\
   \vspace{1cm}}

\begin{center}
\begin{abstract}
In this work we consider the HL-LHC discovery potential in the 3 inverse atto-barn data set for gluinos in the gluino-weakino associated production channel.  We propose a search in the jets plus missing energy channel which exploits kinematic edge features in the reconstructed transverse mass of the gluino. We find that for squark masses in the 2 TeV range we have 5 sigma discovery potential for gluino masses in the range of 2.4 to 3 TeV, competitive with the projections for discovery potential in the gluino pair production channel.
\end{abstract}
\end{center}

\maketitle

\section{Introduction}

The Large Hadron Collider is well into its search for physics beyond
the weak scale. The discovery of a light seemingly fundamental scalar
boson reinforces the urgency of the hierarchy problem. Though Supersymmetry
is a leading paradigm to explain the naturalness of this new particle,
SUSY partners have not yet been discovered. It is assumed that the
first smoking gun signal for Supersymmetry will come in the form of
jets plus missing energy signals from the production of strongly coupled
superpartners, the squarks and gluinos. However, existing searches
already greatly constrain the masses of light-flavored squarks and
gluinos, except in highly mass degenerate scenarios. An estimation
of the gluino/squark lower mass bound in the jets plus $\met$ channel from ATLAS  puts gluino masses
just over 2 TeV, with bounds as high as 1.6 TeV on light generations of squarks \cite{Aaboud:2017vwy}. CMS excludes gluinos decaying to flavorless jets plus $\met$ with masses just above 1.6 TeV, while light-flavored squarks are bound at masses just over 1.3 TeV \cite{Sirunyan:2018vjp}.

The five sigma discovery potential for the colored sparticles is being
approached by already existing limits. For example, upper bounds on the discovery
potential of gluinos in the jets plus $\met$ channel are estimated to be
roughly 2.4 TeV at CMS \cite{CMS}, and around the same at ATLAS \cite{ATLASsen}. In a few scenarios, where decays chain are engineered to
vastly prefer decay through heavy flavored squarks, the 5 sigma discovery
limit on gluino masses is 2.8 TeV \cite{Ulmer:2013csa}. This presents an uncomfortable shadow
scenario for the weak scale physicist, SUSY partners may have masses in the intermediate
TeV range, but yet be undiscoverable with the Large Hadron Collider. In its High Luminosity Run
the LHC will take 3 inverse atto-barns of data. However this represents a sensitivity gain growing only with the square root of luminosity,
therefore some SUSY production modes will simply remain invisible
to searches even with the full  HL-LHC data set. It is then incumbent on the
phenomenologist to switch focus to more rare but spectacular SUSY
production modes which are able to become visible in the high luminosity
data set of LHC. In particular rare events with extremely boosted
states offer an excellent signal to background ratio, and thus hope to offer discovery scenarios to the HL-LHC

In this series of papers we propose to study a set of rare SUSY production
process which increase the discovery potential for superpartners.
This work focuses on the gluino discovery potential. Standard searches
focus on the process of gluino pair production $pp\rightarrow \tilde{g}\tilde{g}$.
However the kinematic threshold in the pair production process severely cuts
off the production mode for heavy gluinos. In addition there are many jets in the events, and large background for the process. Instead we propose
to consider the production of a single gluino
and weakino(either neutralino or chargino),  $pp\rightarrow$
$\tilde{g} \chi_0^{\pm}$, first proposed in reference \cite{Baer:1990rq}. The process yields events with jets plus missing energy however, as weakinos
are expected to be much lighter than gluinos, the kinematic wall for
the gluino mass is significantly relaxed. The events contain a substantial amount of missing energy, and  in addition the relative clean-ness of the events allows us to reconstruct the jets from the gluino decay into kinematic discriminants involving the transverse mass.   Exact production cross section for the process will depend on the admixture of the weakino. Depending on the weakino admixture and weakino mass splittings, there may be a hard leptons in the events which result from the decay of charginos or next to lightest neutralinos. In this work we will thus choose to work in a wino-like weakino scenario.  Wino-like lightest supersymmetric particles(LSPs) present themselves over much of SUSY parameter space as can be seen in classes of models like the PMSSM \cite{Berger:2008cq}, General Gauge Mediated Models \cite{Meade:2008wd}\cite{Carpenter:2008wi}\cite{Carpenter:2008he} \cite{Rajaraman:2009ga}\cite{Carpenter:2017xru}, and extensions of anomaly mediation \cite{Carpenter:2005tz}.  In these scenarios gluino associated production with wino-like charginos is appreciable and the charginos decay to very mass degenerate neutralino LSPs, hence minimizing the number of hard particles in the event.   This work is meant to be a proof of principle of the viability of the gluino-weakino production as a discovery process,   we thus conservatively choose to study a simple inclusive channel with a specialized jets and missing energy analysis. Extending analyses to include other cuts or other weakino scenarios may improve these results even more. By exploiting large missing energies and a kinematic feature in the transverse mass of jets resulting from gluino decay, we will demonstrate that we can provide a good discovery potential for gluino masses in the 2.4 to 3 TeV range.

This paper is organized as follows, Section 2 discusses production
modes and cross sections for gluino-weakino production. Section 3 lays out the SUSY parameter space and event kinematics,
Section 4 describes our cut based analysis. Section 5 gives results for the gluino discovery potential at the HL-LHC. Section 6 concludes.

\begin{figure}[h]
\includegraphics[width=4.50in]{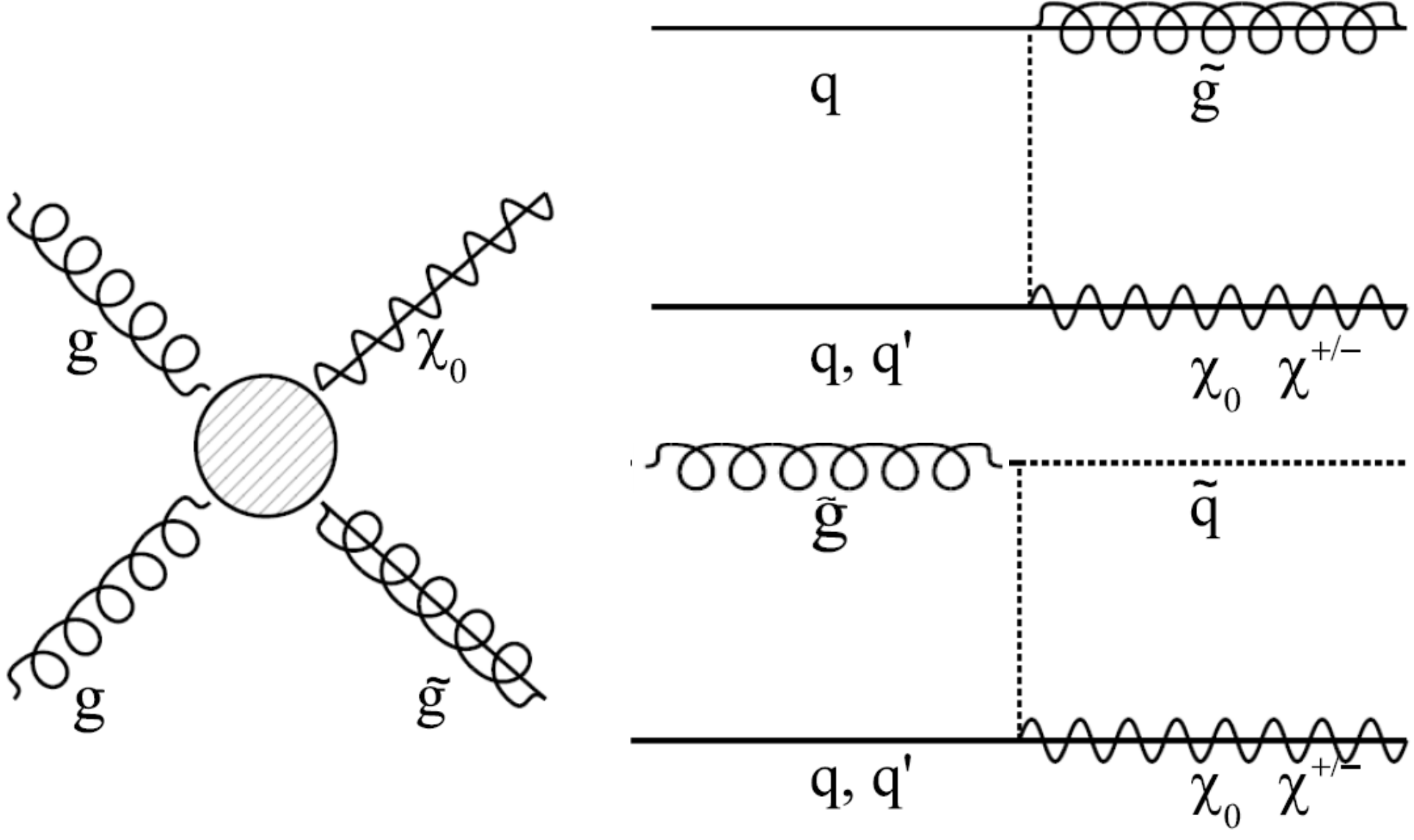}
\caption{Production modes for a single colored MSSM particle in association with a chargino or neutralino.}
\end{figure}

\section{Production Modes}

There are various processes by which a colored SUSY particle may be produced in association with a weakino. We present diagrams of these production mechanisms in figure 1. Gluino-weakino  production follows through a t-channel process through exchange of a virtual squark, the tree level exchange is dominated by the light flavors of squark. The resulting weakino may be charged or neutral. Production of a squark in association  a neutralino or chargino may arise through a quark-gluon fusion process through virtual squark exchange, again the resulting weakino may be charged or neutral.
A loop level process is also possible in which one gluino and one neutralino
are produced through gluon fusion, here all flavors of virtual quarks and squarks run in the loops. We may now explore the relative production cross section of these processes for some typical masses for the MSSM particles using simple benchmark points.

\begin{figure}[h]
\includegraphics[width=4.50in]{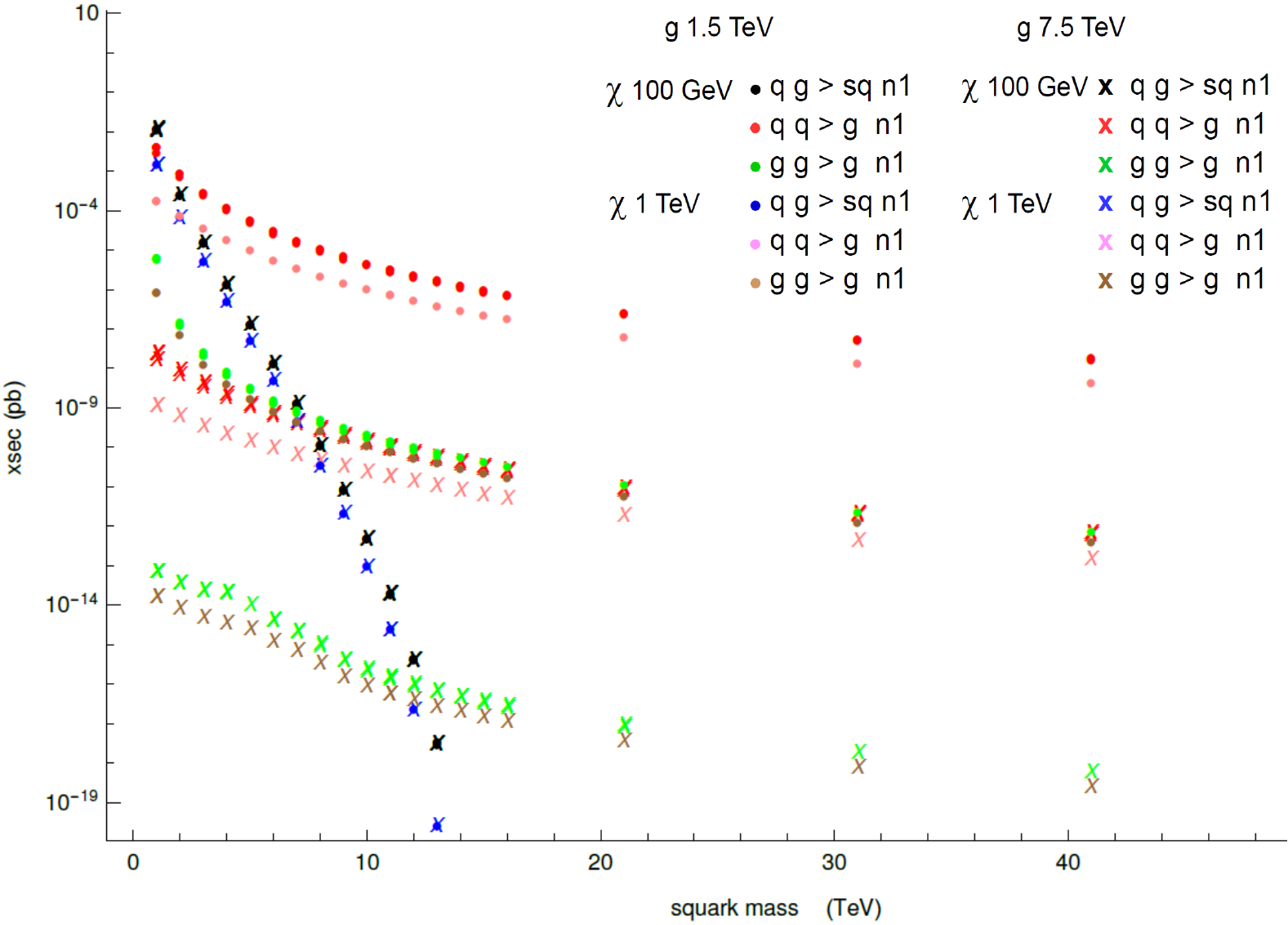}
\caption{Relative production cross sections for colored SUSY particles produced in association with a bino-like neutralino or various benchmark masses.  The production cross sections are plotted as a function of squark masses given in TeV on the x axis.}
\end{figure}

 In figure 2 we show the computed cross sections for the production of squarks or gluinos in association
with the bino-like lightest neutralino.  In order to demonstrate the relative production cross sections of tree level vs loop level processes, the cross sections have been computed using the $Madgraph5@NLO$ \cite{Alwall:2014hca}, and the one loop SUSYQCD model \cite{Degrande:2015vaa}. Since virtual squarks mediate all processes, we have plotted all production cross sections as a function of squark mass, where we have assumed equal masses for all flavors of light flavor squark. Here we have chosen 4 benchmark points;   heavy( m$\chi$=1 TeV) and light (m$\chi=100$ GeV) neutralino mass points, as well as a heavy( $m_g$=7.5 TeV shown as x's) and light($m_g$=1.5 TeV shown as dots) gluino mass points.

We see that in the squark-neutralino production mode, the production
cross section falls off very sharply as the squark mass approaches
the kinematic threshold of 14 TeV exactly as we expect.
In examining the  neutralino-gluino production we see that increasing
the neutralino mass from effectively zero to 1 TeV produces an order of
magnitude difference in cross section. We also note that for large
values of squark mass, above the average center of mass energy of LHC collisions, the  gluino squark production process is
well approximated by a dimension 6 effective operator proportional
to $\frac{1}{m_{sq}^{2}}$. Tree level gluino neutralino production is the dominant process over a large range of squark masses, and remains above the a femto-barn for squark masses of a few TeV.  The loop-level gluino neutralino production from
gluon fusion is much smaller than the production from quark fusion except in the case of very light squarks, and falls rapidly with the squark mass. We therefore neglect loop level production in this work.

\section{Event kinematics and SUSY parameter space}

Our study of gluino-weakino production relies on four parameters, the weakino and gluino masses, the weakino mixing content, and the squark masses. In this analysis, we will fix the weakino content by demanding that the lightest neutralino be purely wino-like. In the wino-like the case, the LSP neutralino will be accompanied by an almost mass degenerate pair of charginos, split from the LSP mass by an amount which must be more than a pion mass \cite{Chen:1999yf}. In our simplified model only the squarks, gluinos and wino-like weakinos will be light with all other SUSY masses including sleptons, the bino, and  Higgsinos very large, the choice of heavy Higgisnos and bino is theoretically consistent with a wino-like LSP. The gluinos are produced in association with both light charginos and neutralinos. In our studies we have fixed the neutralino mass to be 100 GeV. This is phenomenologically viable, as in searches with very small mass splittings between the neutralino and chargino, light neutralino masses are mostly unconstrained by current disappearing track or mono-boson searches \cite{Aaboud:2017mpt} \cite{Anandakrishnan:2014exa} \cite{Baer:2014cua}. Disappearing track searches will have sensitivity to order 100 GeV wino scenario for a narrow range of mass splittings in the full HL-LHC run \cite{ATLAS} and mono-boson searches may constrain some parameter space. Having fixed the mass of neutralinos and charginos we may now consider the  kinematics of our signal events as a function of gluino and squark masses.

In our simplified SUSY spectrum the decay modes of the gluinos will be $\tilde{g}\rightarrow \overline{q}\tilde{q} \rightarrow \overline{q}q\chi_{0}$ and $\tilde{g}\rightarrow \overline{q}\tilde{q} \rightarrow \overline{q}q^{'}\chi^{\pm}$. Note here the decay of the gluino may proceed through either on or off shell intermediate squarks.  The chargino decay will proceed through an off shell W to the neutralino and soft products. The signal events will thus be amenable to an inclusive jets plus missing energy search.

One main characteristic of these events is a substantial amount of missing
energy. Recall that the definition of missing transverse energy is that it is
the negative vector sum of the visible transverse momentum, $\met = -\Sigma \overline{p_T}$. We may examine the characteristic distributions of missing energy using collider simulation techniques.  We generated samples of signal events using Madgraph \cite{Alwall:2014hca}, events were decayed with Madspin\cite{Artoisenet:2012st}, showered  using pythia \cite{Sjostrand:2014zea}, and run through the pgs detector simulator \cite{pgs}. Events were generated  applying a 300 GeV generator level $\met$ cut.

The upper right hand corner of figure 3 shows the $\met$ distribution of 30000 signal events compared to the Zjj background for a benchmark point with $m_{\tilde{g}}$ 2.6 GeV and $m_{\tilde{q}}$ 1.6 TeV. This benchmark point involves gluino decay through an on-shell squark. In the rest frame of the gluino we may compute the squark's momentum, $p_{\tilde{q}}$, which is approximately $\left({m_{\tilde{g}}}^2 - {m_{\tilde{q}}}^2\right)$/$ 2m_{\tilde{g}}$. The squark subsequently decays to a nearly massless quark and much lighter neutralino/chargino leaving the visible momentum in the event substantially unbalanced. The maximum $\met$ occurs when the squark is ejected in the fully transverse direction, in which case we would expect the characteristic missing energy to be around .8 TeV in the gluino's rest frame for our benchmark point. The characteristic missing energy of an event increases with increasing gluino mass as expected and also increases with increasing weakino mass.

In figure 3 we have also plotted the distribution of the dijet invariant mass of our events, in the upper left corner. In out figure,the invariant mass is computed by taking the invariant mass of the two leading hard jets, which are overwhelmingly likely to come from the decay of the gluino.  For our benchmark point with ${m_{\tilde{q}}} >> m_{\chi 0}$ the maximum invariant mass should be approximately $ p_q\sqrt{(p_q^2+{m_{\tilde{q}}}^2)}+p_q^2$,  about 2 TeV for out benchmark point.  We see this estimate agrees with our plot.

In order to characterize the kinematics of the event we may construct the transverse  mass of the leading di-jets.   The standard transverse mass is given by $m_{T0}^2= (\Sigma {E_T})^2 - (\Sigma \overline{p_T})^2$. For our benchmark point,  the minimum transverse mass occurs when the gluino decay ejects the initial squark in the longitudinal direction. The transverse mass of the dijet system will depend on the azimuthal angle of the initial squark, so we expect the  transverse mass to then be distributed uniformly until the maximum occurs when the gluino decay ejects the squark in the purely transverse direction. For our benchmark points we may compute this at bit under 2 TeV.  In the lower left of figure 3 we show the distribution of $m_{T0}$ consistent with our expectations.

We see that with in our $m_{T0}$ distribution, we have a significant number of events populating the low transverse mass region.  This is less efficient for distinguishing signal from background as there is significant overlap in the low transverse mass region. Therefore, in analyzing our events, we also use a generalization of the 'inclusive' transverse mass which we define as $m_{Ti}^2= (m_{iv}^2 + (\Sigma \overline{p_T})^2)$.  We see that with this definition,  the inclusive transverse mass of our leading dijet system will be guaranteed to be larger than the invariant mass. We thus expect that, compared to the di-jet invariant mass distribution, the $m_{Ti}$ distribution should have more events at high values.  We expect a maximum of the inclusive transverse mass again when the gluino decay ejects the squark in the transverse direction. Our rough calculation for on-shell squarks yields a maximum value  $ p_q\sqrt{(p_q^2+{m_{\tilde{q}}}^2)}+p_q^2 +p_q^2/4$. With the characteristic value of this distribution being a bit above that of the invariant mass distribution, we  also expect the $m_{Ti}$ distribution to favor higher values (and have fewer events at low value) as compared to the  $m_{T0}$ distribution. The leading di-jet $m_{Ti}$ distribution is plotted in the lower right of figure 3. The distribution shape conforms our expectations from simple kinematics.


\begin{figure}[h]
\includegraphics[width=3.2in]{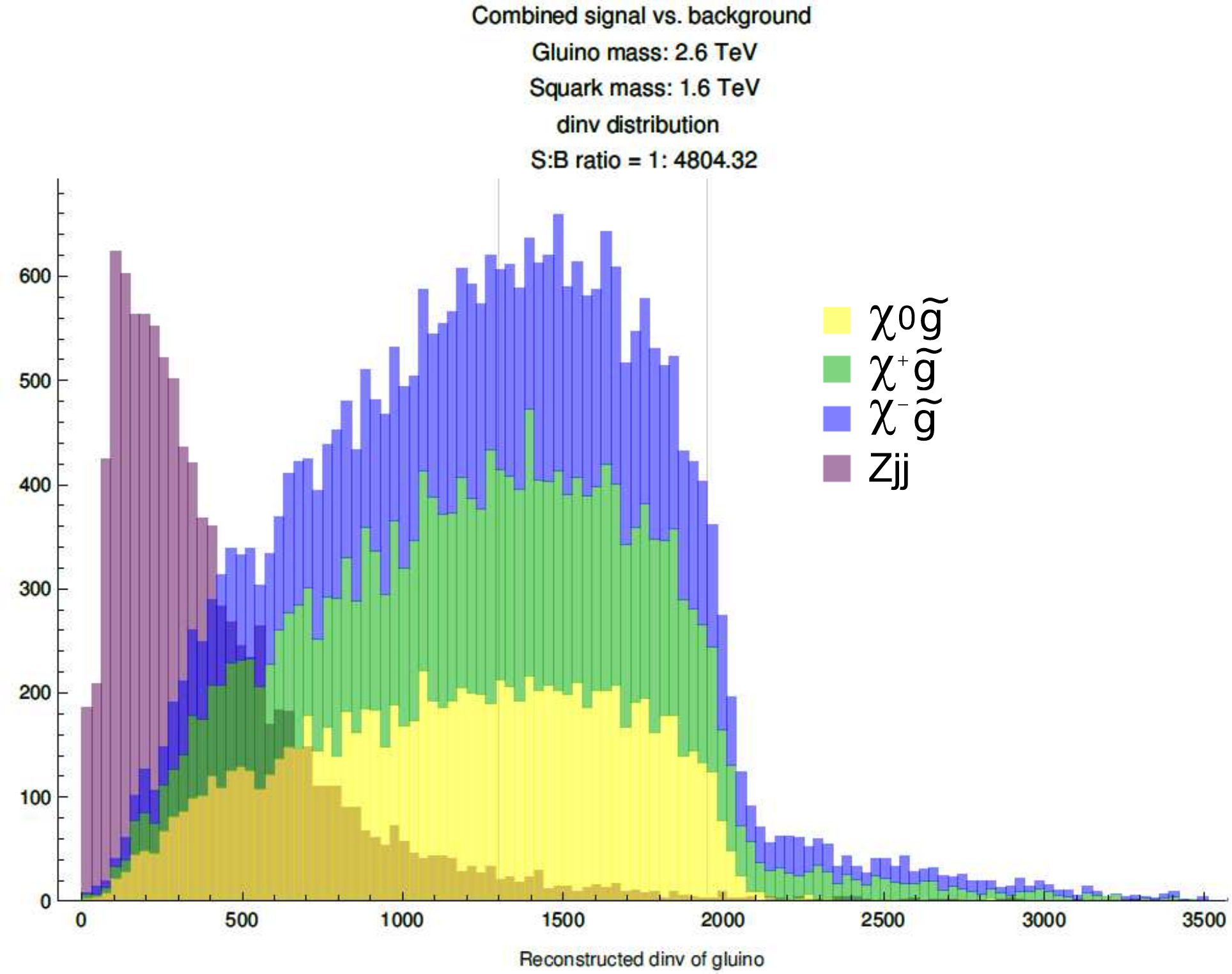}
\includegraphics[width=3.2in]{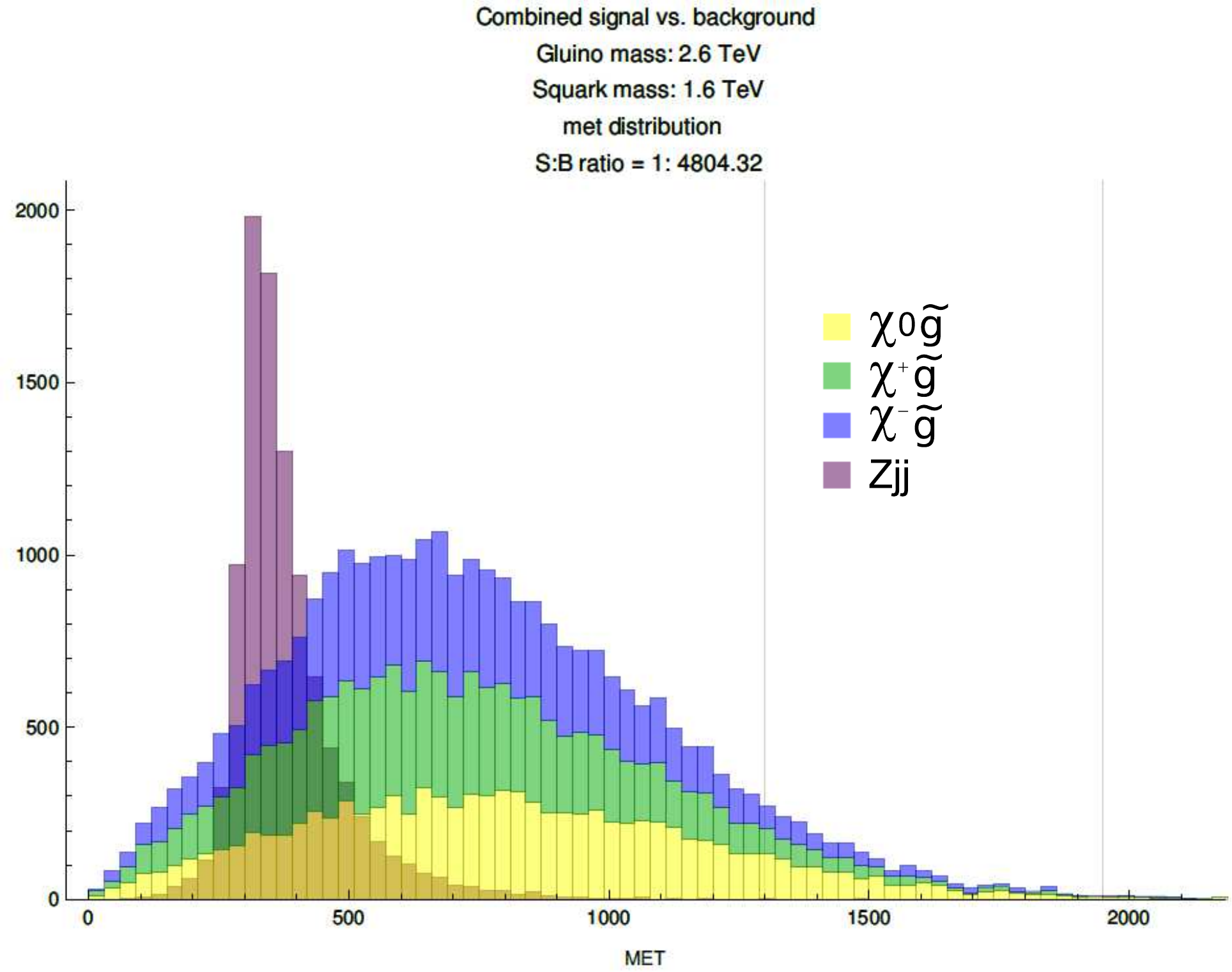}\\
\includegraphics[width=3.2in]{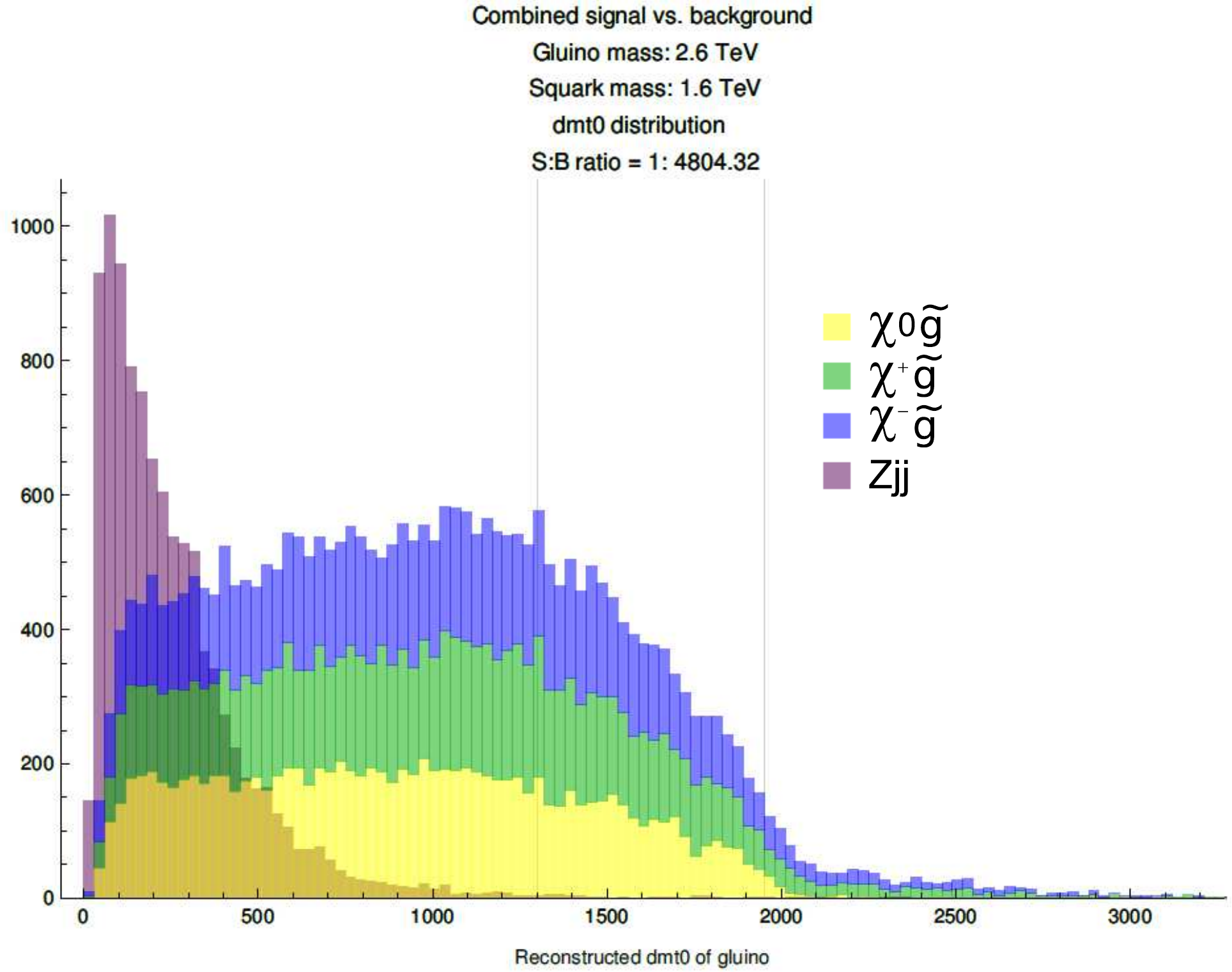}
\includegraphics[width=3.2in]{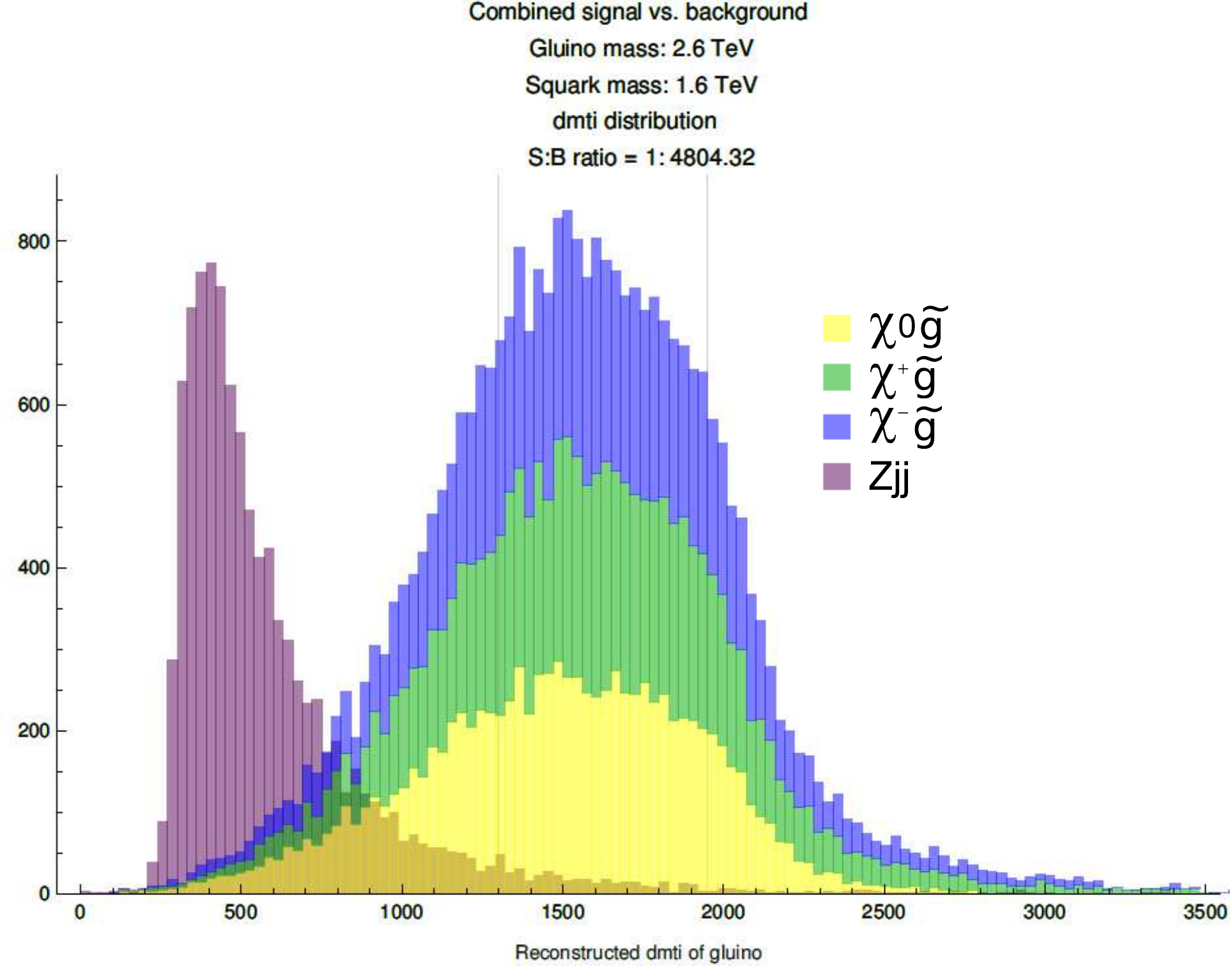}\\
\caption{Histograms giving the distribution of various kinematic discriminants in signal and background events of sample size 10000.  The upper left plot gives the di-jet invariant mass distribution of events; the upper right gives the $\met$ distribution of events; the lower left shows the di-jet $m_{T0}$ distrbution, and the lower right shows the di-jet $m_{Ti}$ distribution.}
\label{fig:models}
\end{figure}

\section{Cut based analysis}

Having discussed the signal kinematics we now describe our cut based jets plus missing energy analysis which has been taylored for this signal.  We note that in the production of signal events we have scaled the tree level cross section  predictions from Madgraph with a modest k factor of 1.3, which is consistent with next to leading order computations for this process \cite{Berger:1999mc}\cite{Fuks:2016vdc}. The main background for this process consists of Z + jets production in which the Z boson decays to neutrinos; $q \overline{q} \rightarrow Z jj \rightarrow \met+ j j$. In testing the possible t-tbar background as a source of jets plus missing energy events, we found the number of events passing cuts to be negligible in our analysis compared to the main Z+jets process. Background events were also created with consistent 300 GeV generator level missing energy cuts. Background events were generated  using Madgraph, showered with pythia, and passed through the pgs detector simulator.  The kinematic distribution of missing energy, invariant di-jet mass, $m_{T0}$ and $m_{Ti}$ are given along with the signal in figure 3.  We see in the distribution plots  that  resulting $\met$ distribution for the background is peaked at small values and swiftly falling with increasing missing energy. The transverse  and invariant mass distributions are also peaked at low value and fall off very quickly at high values. In order to separate signal from background in our analysis we therefore consider the following cuts,

\begin{table}
\begin{tabular}{|c|c|c|c|c|}
\hline
$m_{\tilde{g}}$ 2.2 TeV & & & & \tabularnewline
\hline
cut & $\tilde{g}\chi_0$ & $\tilde{g}\chi^{-}$ &  $\tilde{g}\chi^{+}$ & Z j j \tabularnewline
\hline
\hline
 none & 10000 & 10000 &  10000 & 10000 \tabularnewline
\hline
 inclusive $m_{T0}$ & 1511 & 2417 & 2509 & 94 \tabularnewline
\hline
500 GeV $\met$ & 1048 & 1592 & 1645 & 11 \tabularnewline
\hline
\hline
$m_{\tilde{g}}$ 1.0 TeV & & & & \tabularnewline
\hline
cut & $\tilde{g}\chi_0$ & $\tilde{g}\chi^{-}$ &  $\tilde{g}\chi^{+}$ & Z j j \tabularnewline
\hline
\hline
 none & 10000 & 10000 &  10000 & 10000 \tabularnewline
\hline
 inclusive $m_{T0}$ & 2269 & 4091 & 4259 & 1091 \tabularnewline
\hline
500 GeV $\met$ & 796 & 583 & 532 & 124 \tabularnewline
\hline
\end{tabular}
\caption{Cut flow for signal and the main Z jj background for 2 benchmark points with 1.6 TeV quarks. The transverse mass $m_{T0}$ is reconstructed using the exclusive di-jet method. To demonstrate the change in efficiencies as the transverse mass window shifts with gluino mass, the top benchmark point gives cut flow for a  2.2 TeV gluino while the bottom benchmark point gives cut flow for a  for 1 TeV gluino.}
\end{table}

\begin{itemize}
\item Events must contain at least 2 jets in the central region of the detector $\eta <2.5$
\item Jets must have $p_T > 20 GeV$
\item Events must have $\met \ge$ 500 GeV
\end{itemize}
From this point we now test four possible cut flows. We choose to construct and cut around one of two transverse masses, either  $m_{T0}$ or $m_{Ti}$. We choose to construct these transverse masses one of two possible ways, using either an exclusive di-jet or an inclusive all-jet method. First we describe the exclusive method,

\begin{itemize}
\item A dijet transverse mass discriminant $m_{T} = m_{T0}$ or $m_{Ti}$ is constructed using only the two leading jets in the event
\item the chosen $m_{T}$ must fall in a kinematic window which varies with hypothesized gluino mass $M_g - \Delta < m_{T} < M_g + \Delta$ where  $\Delta$
is .5$M_g$
\end{itemize}

Next we describe the inclusive all-jet method,

\begin{itemize}
\item  A  transverse mass discriminant $m_{T} = m_{T0}$ or $m_{Ti}$ is constructed using all viable jets
\item the chosen $m_{T}$ must fall in a kinematic window which varies with hypothesized gluino mass $M_g - \Delta < m_{T} < M_g + \Delta$ where  $\Delta$
is .5$M_g$
\end{itemize}

To show the difference in sigmal and background distributions using our four cut-flow techniques, we have created figure 4. In this figure, we have chosen a benchmark point with gluino mass of 2.6 TeV and squark mass of 1.6 TeV.  We show scatter plots for 10000 signal events and 10000 background events in the missing energy-transverse mass plane. The upper left plot shows $\met$ vs $m_{T0}$ using the exclusive dijet method, the upper right shows  $\met$ vs $m_{Ti}$ using the exclusive di-jet method. The lower left plot shows $\met$ vs $m_{T0}$ using the inclusive all-jet method, the upper right shows  $\met$ vs $m_{Ti}$ using the inclusive all-jet method.  We see that the inclusive all-jet methods have the predictable effect of smearing out the events. We also note the events take a characteristic distribution in the missing energy-transverse mass plane, which would be an interesting subject for efforts to further optimize this analysis.

In table 1 we present a sample cut-flow for signal and background events for two possible benchmark points. We have used the exclusive di-jet method to construct the $m_{T0}$ transverse mass  for the cut flow in this table.  To show how the search efficiencies depend on the gluino mass we show cut-flows for a benchmark point with squark mass 1.6 and  gluino mass 2.2 TeV, and another benchmark point with squark mass 1.6 TeV and gluino mass 1 TeV.  We see from the cut-flow that as we raise the gluino mass we also raise threshold to make it into the transverse mass window. The high transverse mass threshold ensures  that the ratio of signal to background efficiency decreases drastically for appreciable gluino masses.

\begin{figure}
\includegraphics[width=3.2in]{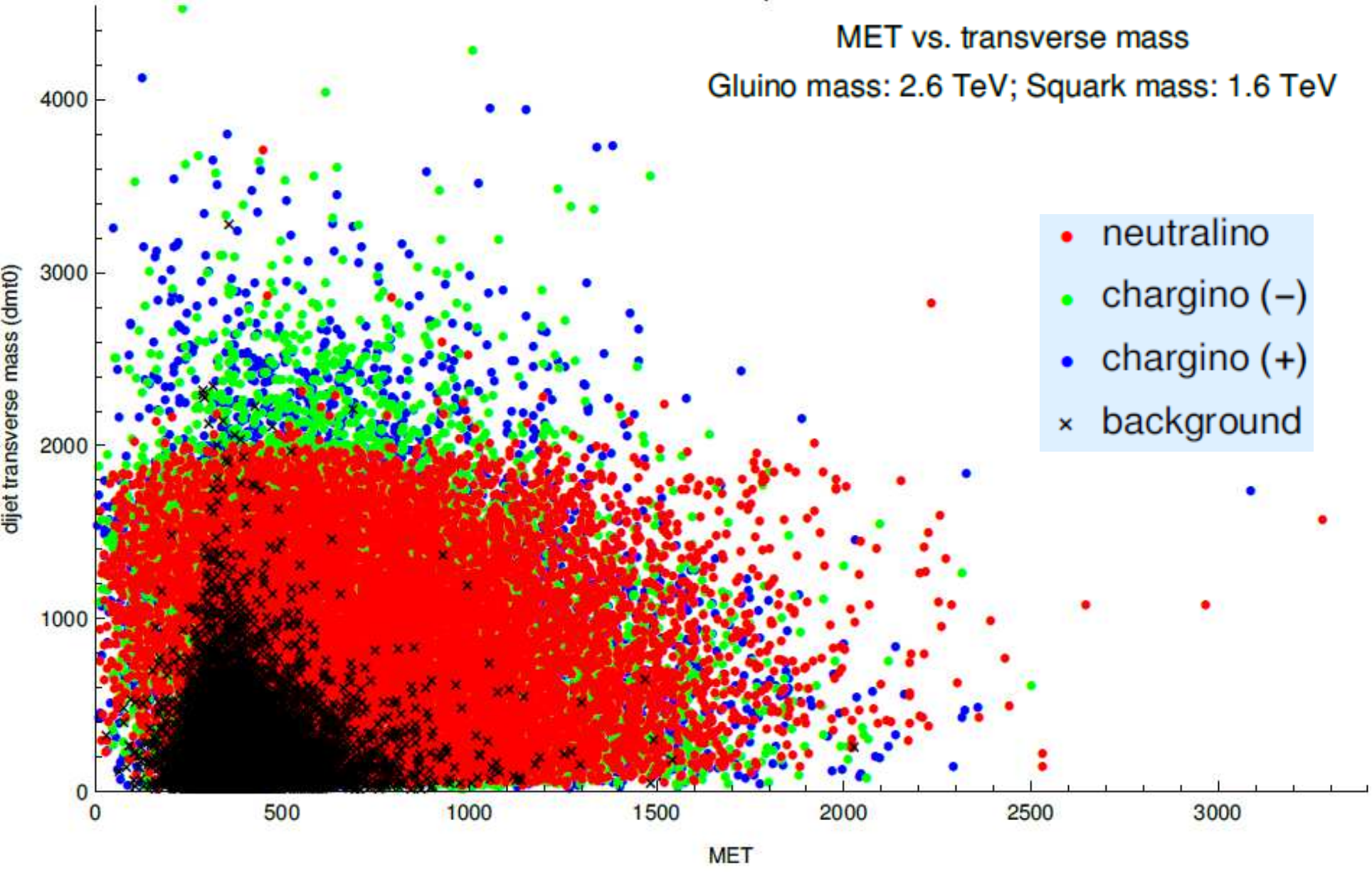}
\includegraphics[width=3.2in]{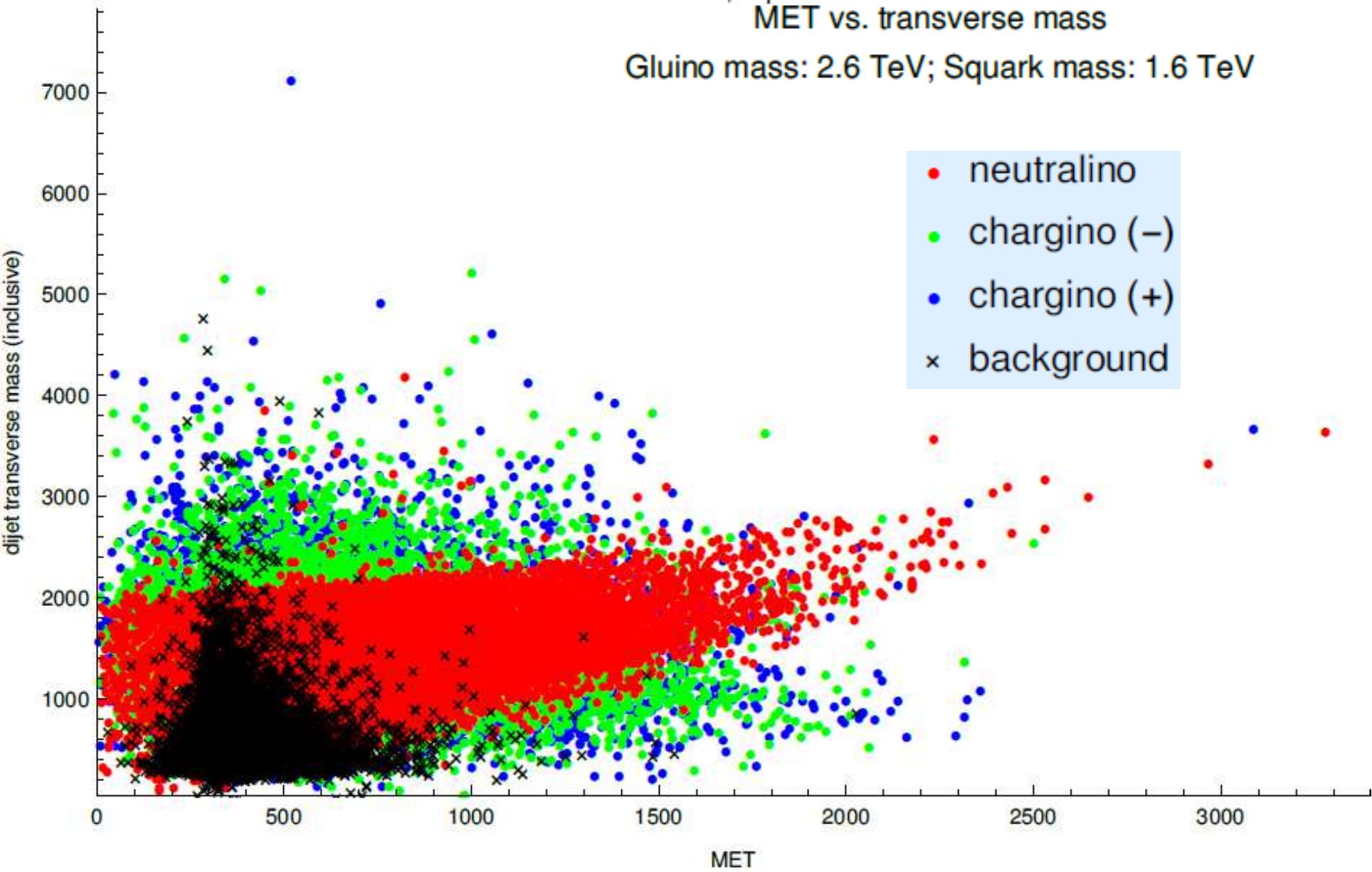}\\
\includegraphics[width=3.2in]{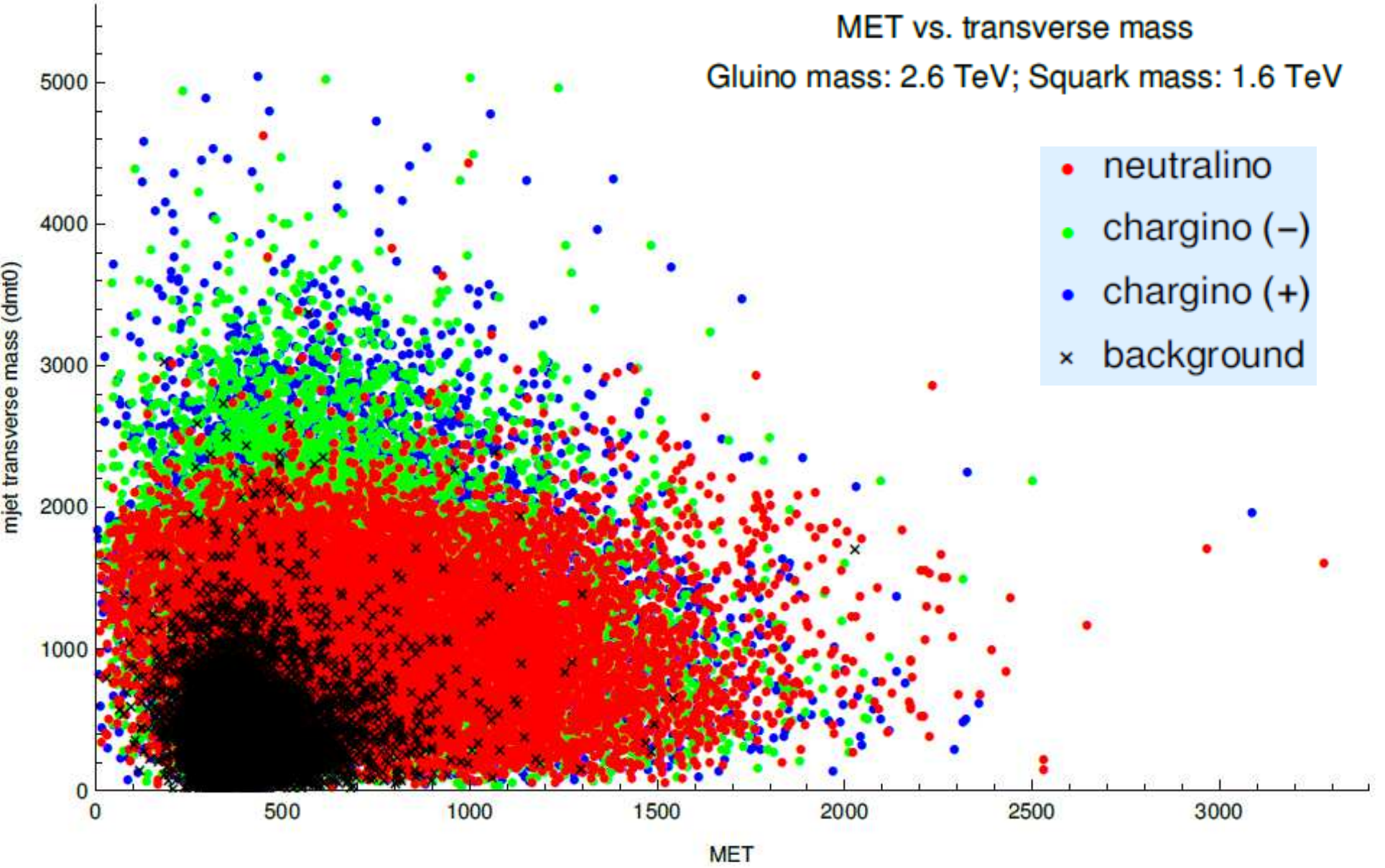}
\includegraphics[width=3.2in]{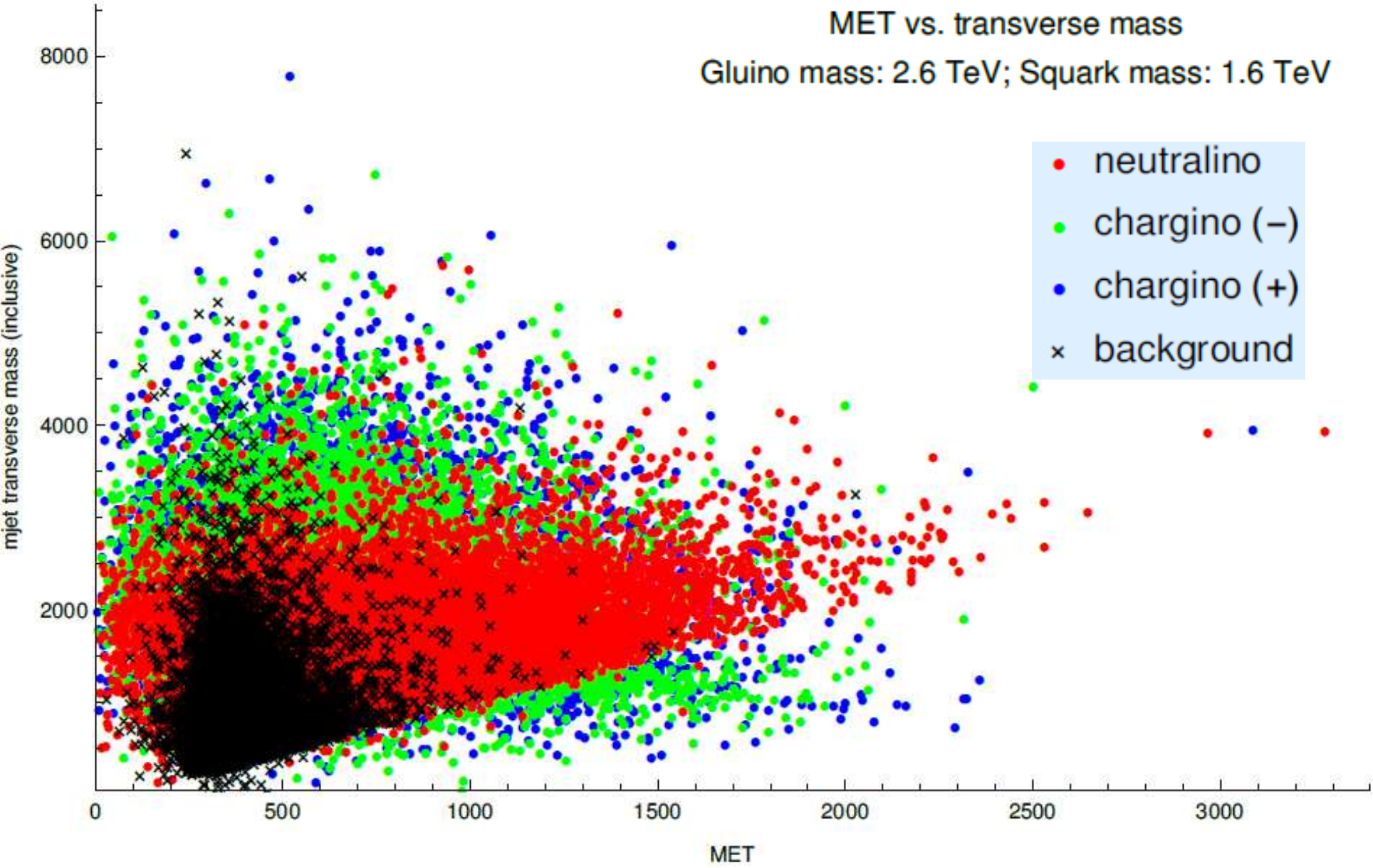}\\
\caption{Scatter plot of missing transverse energy vs various invariant masses in signal and background events. The black dots are background, the red blue and green dots show events with gluinos produced in association with $\chi_0, \chi^{+}$ and $\chi^{-}$ respectively. The upper plots show the exclusive di-jet  cut method with the left giving the distribution of $m_{T0}$ and the right giving $m_{Ti}$. The lower plots show the inclusive all-jet method with  the left giving the distribution of $m_{T0}$ and the right giving $m_{Ti}$.}
\label{fig:models}
\end{figure}

\section{Results}

\begin{figure}
\includegraphics[width=4.3in]{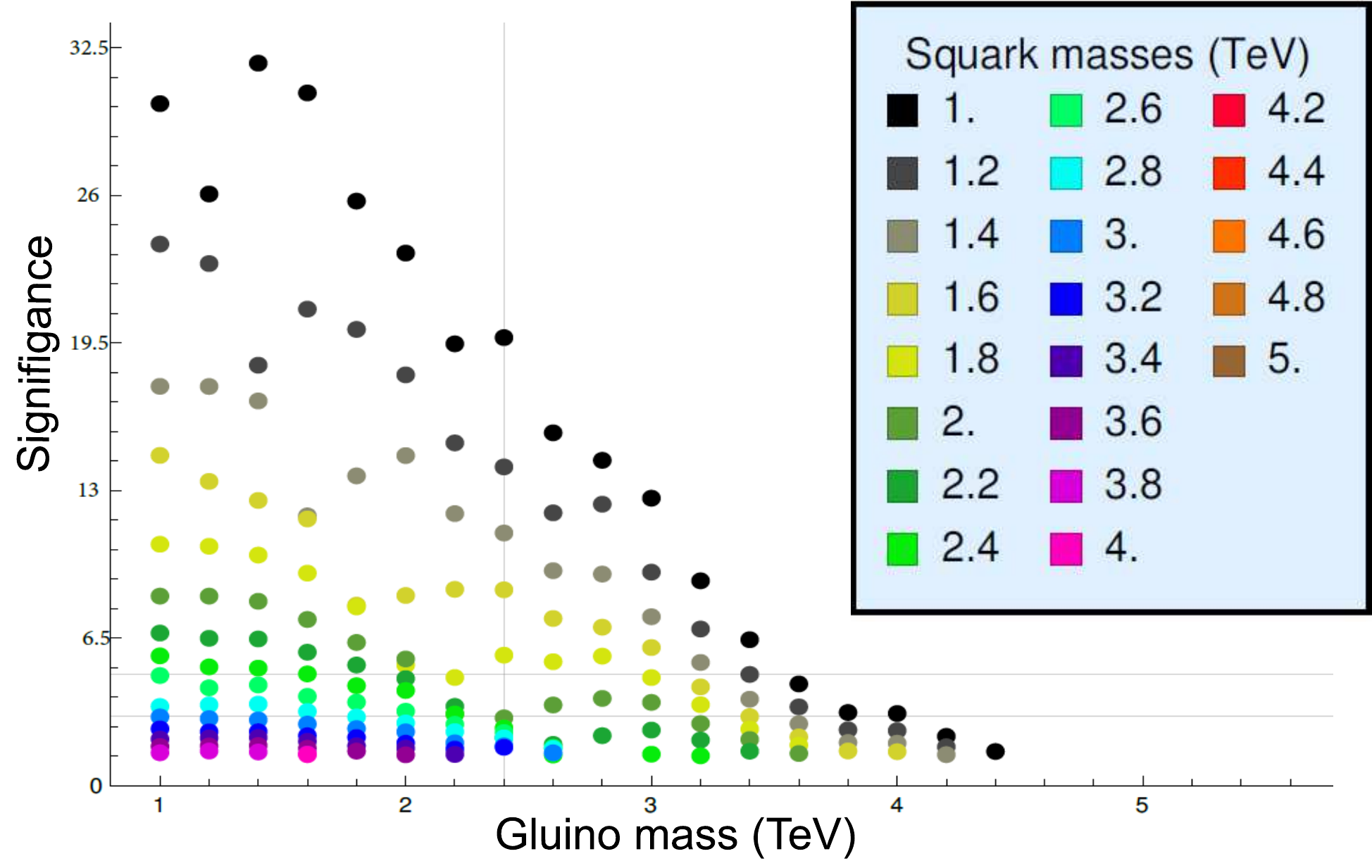} \includegraphics[width=4.3in]{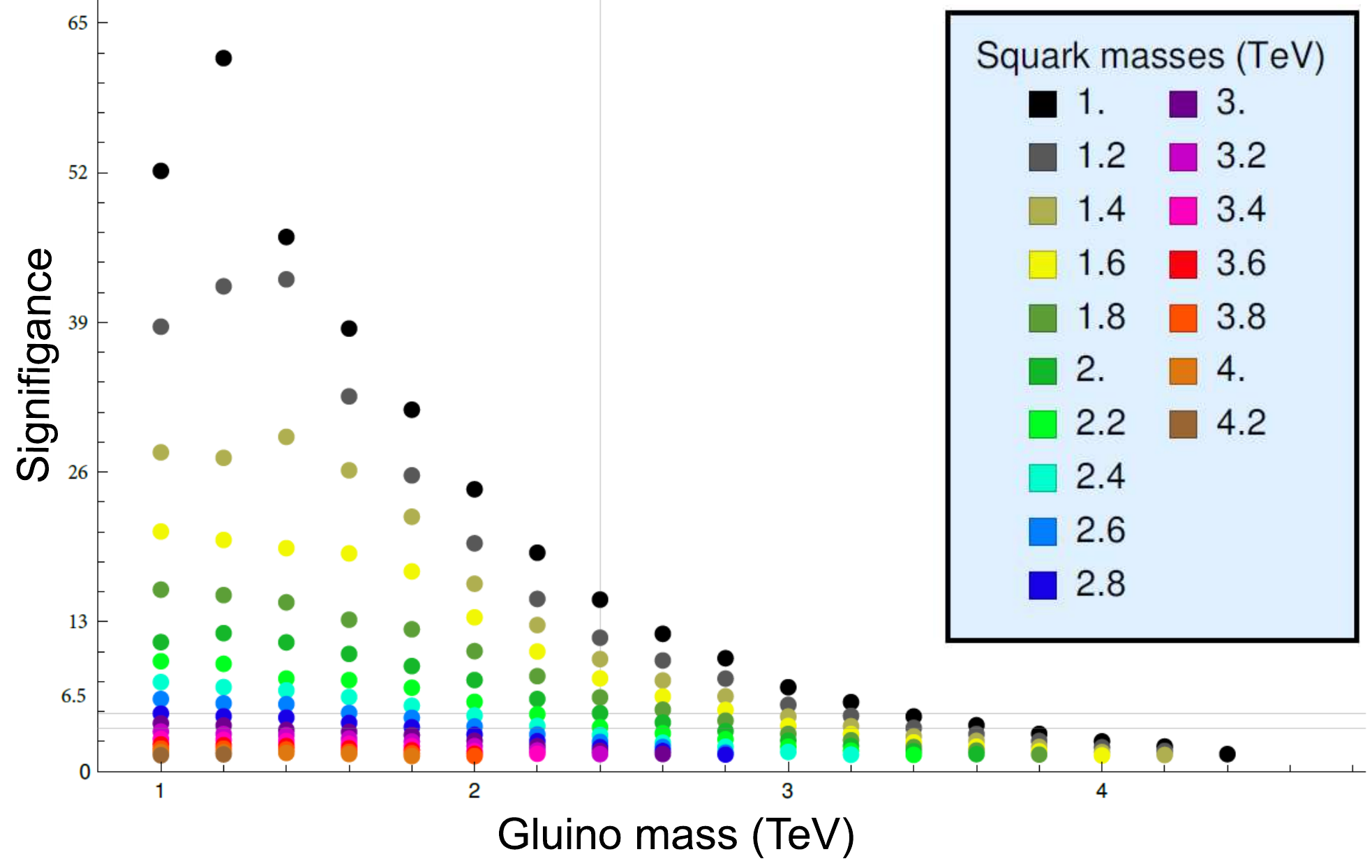}\\
\caption{Significance for gluino-weakino production vs gluino mass for various squark masses. The upper plot gives significances for the search which uses the di-jet $m_{T0}$ tansverse mass discriminant. The lower plot gives the significance for the search which uses the di-jet $m_{Ti}$ tansverse mass discriminant.}
\label{fig:models}
\end{figure}

We will now construct the discovery potential for gluinos for the 3 inverse atto-barn run of the HL-LHC. Using conventions for low background statistics we may define the signal significance, $S$.  With Sg as the number of signal events and B the number of background events $S$=Sg/$\sqrt{B}$. The conventional threshold for discovery is to take $S$=5 for discovery and $S$= 3 for sensitivity \cite{Baer:2009dn}.

In figure 5 we have plotted significance vs. gluino masses for various squark masses in the gluino-weakino associated production channel.  We have added horizontal lines to indicate the  significance thresholds of 5 for discovery and 3 for sensitivity. In addition, we have added a vertical line at a gluino mass of 2.4 GeV for comparison to current stated discovery potential. In the upper plot we have shown results using our exclusive di-jet search, constructing the transverse mass $m_{T0}$. In the lower plot we show results using our exclusive di-jet search constructing the transverse mass $m_{Ti}$. For comparison, we give a plot using the inclusive method in the Appendix.

We see from figure 5 that the gluino discovery potentials in our scenario depend heavily on squark masses. ATLAS sets the current
limits for squark masses decaying 100 percent of the time via $ \tilde{q}\rightarrow q \chi_0$  to jets plus missing energy at 1550GeV for 100 GeV neutralino masses \cite{Aaboud:2017vwy}. CMS places limits in this scenario of 1325GeV \cite{Sirunyan:2018vjp}. ATLAS searches in the 1-lepton final state bound squark masses to be at least 1200 GeV \cite{Aaboud:2017bac}, but this result involves a squark decaying through on-shell W bosons to hard leptons $\tilde{q}\rightarrow q \chi^{\pm}\rightarrow qW\chi_0$.  In our scenario the squark has a significant branching fraction into charginos that are highly mass degenerate with the neutralino LSP, it is then a question as to what squark lower mass limits are in this scenario. The 0-lepton search limits present to us a conservative a conservative choice of lower squark mass bounds.

We can see that in our search constructing  $m_{T0}$ from exclusive di-jets,  we find a 5 sigma discovery potential for gluinos with masses of 3.1 TeV for 1.6 TeV squarks. For 1.6 TeV squark masses we find a 3 sigma sensitivity potential for gluinos of masses 3.4 TeV. One will notice a  feature in this sensitivity plot that appears once the possibility of decay through an on-shell squark becomes kinematically possible for the gluino, improving the search  efficiency. In the lower plot, using the $m_{Ti}$ discriminant constructed from exclusive di-jets, we can see that for squark masses of 2.2 TeV, we have a 5 sigma discovery potential for gluinos of 2.2 TeV with  a 3 sigma sensitivity potential for gluinos of about 2.5 TeV. Our results are competitive with current projections and may raise the discovery potential for gluino masses above 2.4 in the case of lighter squark masses.

\section{Conclusions}

We have demonstrated the 5 sigma discovery potential for the 3 inverse atto-barn HL-LHC may extend to gluinos masses in the 2.4 to 3 TeV range by studying the  gluino-weakino production channel. The discovery potential in this case is competitive
with that of the standard gluino pair production channel. The resultant discovery potential
comes despite smaller a production cross section than the gluino pair production process. However, the stand-out
kinematics of the gluino recoiling off of a light weakino allows our analysis large missing energy cut, and very substantial di-jet transverse mass cut. The resulting search has a low background rate and ensures the gluinos are discoverable.

This work offers an existence proof that the gluino discovery
potential may be substantial in the gluino-weakino channel by employing a very basic cut based analysis.  We thus note that with optimization and
improvements of the gluino-weakino analysis, it is possible the discoverable gluino mass threshold may be raised even more. We will discuss some opportunities for expanded searches. As was mentioned before, a more sophisticated kinematic cut may be engineered to take into account the relation of missing energy to the transverse mass in the signal events.  In addition, more sophisticated cuts may be made to take into account the  shape of the distributions of the kinematic variables. These edge effects have been discussed in proposed searches for supersymmetric particles   \cite{Allanach:2010pp}and exotics such as Dark Matter \cite{Abercrombie:2015wmb}. Further, alternate regions of SUSY parameter space may be studied. One example is regions of SUSY parameter space where squark
decay channels may be altered to include intermediate states like
second to lightest neutralinos or on-shell vector bosons. This may add hard leptons to the events.
In addition, lower mass limit on squarks in these scenarios are
more loose, in which case the gluino production cross section may be increased.

Another possibility is to consider models which split the masses of squark flavors. We
have operated under the assumption that 4 flavors of squark are mass
degenerate. This gives the toughest lower mass bounds to squarks which
decay to quark plus neutralino. Splitting squark flavors may
relax the lower mass bound on squarks and have interesting effects on production cross sections of the gluino weakino pair due to differing quark pdfs.

Finally, in the wino or Higgsino like weakino scenarios, the mass splitting
of the chargino and LSP might be adjusted to ensure that the chargino
resultant in the gluino decay lives an intermediate amount of time and appears as a disappearing track. An
additional disappearing track and two hard jets may produce a great
discovery scenario for this process.

\section{Appendix}
For comparison we give a plot of signal significant vs. gluino mass for a range of squark masses in a search using the all jet inclusive $m_{T0}$ discriminant.

\begin{figure}
\includegraphics[width=4.3in]{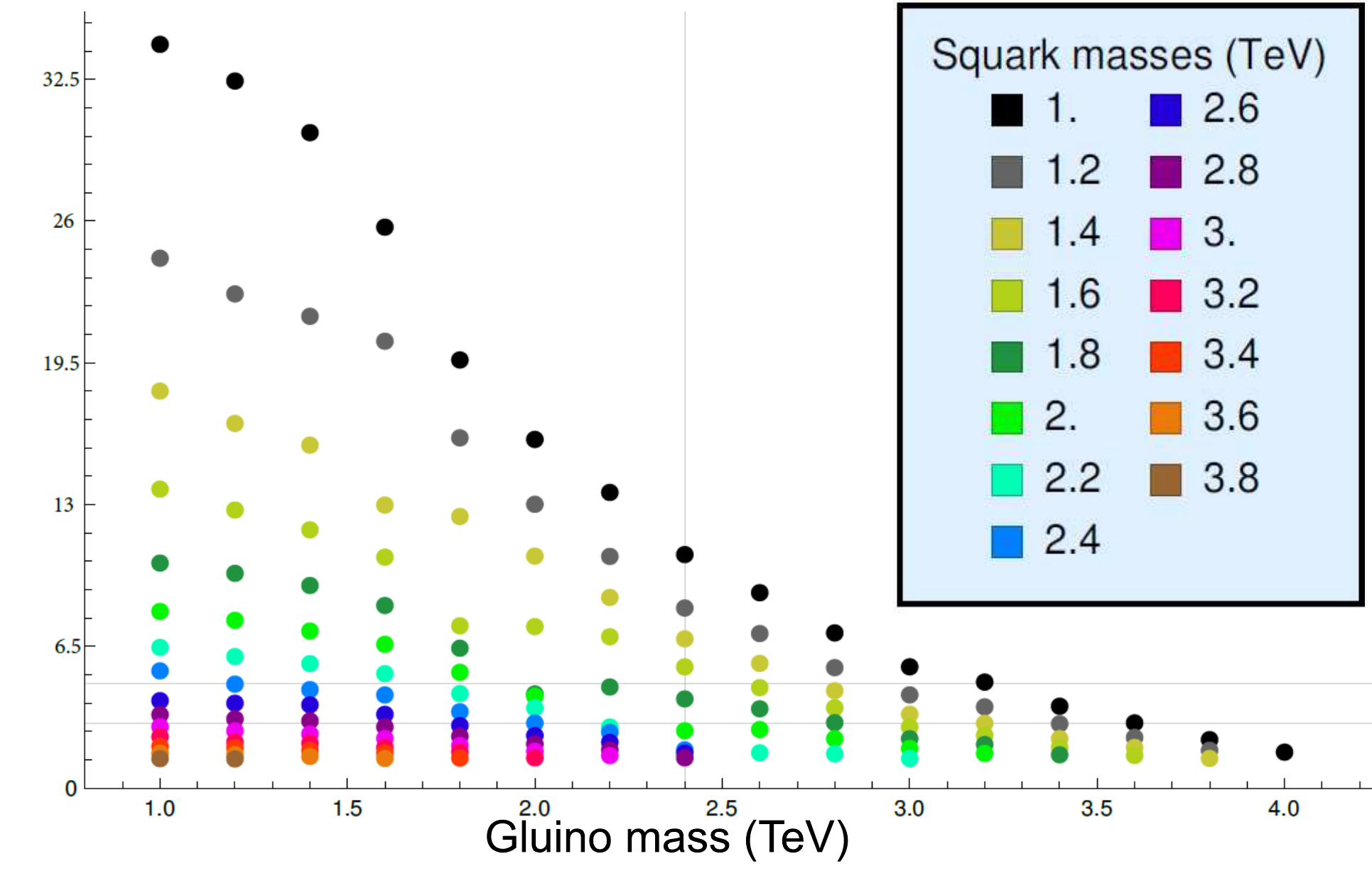}
\caption{Significance for gluino-weakino production vs gluino mass for various squark masses. Significances  are given for the search which uses the all-jet $m_{T0}$ transverse mass discriminant. }
\label{fig:models}
\end{figure}

\section{Acknowledgements}
This work was supported in part by the US department of Energy grant DE- SC0011726. We thank Antonio Boveia for discussions.

\end{document}